\documentclass[preprint,amsmath,amssymb,superscriptaddress,floatfix]{revtex4-2}
\usepackage{printlen}
\usepackage{lineno} 
\usepackage[T1]{fontenc}
\usepackage[english]{babel}
\usepackage{graphicx}
\usepackage{amsmath,amssymb,mathrsfs}
\usepackage[dvipsnames]{xcolor}
\usepackage[caption=false]{subfig}
\usepackage{float}
\usepackage{numprint}
\usepackage{siunitx}
\usepackage{todonotes}
\usetikzlibrary{matrix}
\usepackage{soul}
\usepackage{array}
\usepackage{hyperref}
\usepackage{nameref}
\usepackage{booktabs}
\usepackage[referable]{threeparttablex}
\usepackage{longtable}
\usepackage{dcolumn}
\newcolumntype{L}{D{.}{.}{2,5}}
\usepackage{adjustbox}
\usepackage{tabularx}
\usepackage{multirow}
\usepackage{makecell}

\usepackage[version=4]{mhchem}

\usepackage{siunitx}

\usepackage{tikz-3dplot}
\usetikzlibrary{decorations.markings,arrows,arrows.meta, decorations.pathmorphing,backgrounds, positioning, fit, shapes.geometric,shapes,bending,shadows,shadings,shapes.misc,spy}
\usetikzlibrary{calc}
\usetikzlibrary{math}

\begin{document}
\title{Ferroelectric Properties and Topological Polar Textures of PbTiO$_3$ from a Second-Principles Open-Source Interatomic Potential}
\author{Louis Bastogne}
\email{louis.bastogne@uliege.be}
\affiliation{Theoretical Materials Physics, Q-MAT, Université de Liège, B-4000 Sart-Tilman, Belgium}%
\author{Philippe Ghosez}
\email{Philippe.Ghosez@uliege.be} %
\affiliation{Theoretical Materials Physics, Q-MAT, Université de Liège, B-4000 Sart-Tilman, Belgium}%
\date{\today}
\begin{abstract} 
We introduce an open-source, fully atomistic second-principles interatomic potential for lead titanate (PbTiO$_{\mathrm{3}}$), a benchmark ferroelectric material known for its strong polarization and high-temperature phase transitions. While density functional theory excels at capturing atomic-scale behavior, it remains computationally prohibitive for large-scale simulations required to explore complex phenomena. Our model addresses this limitation by accurately reproducing key properties of PbTiO$_{\mathrm{3}}$, including domain wall dynamics and different topological textures formation, which are known as key features to next-generation memory and energy-efficient technologies. Validated against DFT data, the model remains predictive across a wide range of conditions. It offers an accessible and efficient framework for high-accuracy large-scale simulations, allowing deeper insights into PbTiO$_{\mathrm{3}}$ and its potential applications.
\end{abstract}

\maketitle

\newpage


\section{Introduction}

Lead titanate (PbTiO$_{\mathrm{3}}$) is a cornerstone material in the ferroelectric (FE) community, well known for its large spontaneous polarization~\cite{nelmes1985crystal} and high-temperature ferroelectric-to-paraelectric phase transition~\cite{shirane1951phase}. These attributes underpin its widespread use in applications such as transducers and actuators~\cite{zhang2013relaxor}. They also establish PbTiO$_{\mathrm{3}}$ as a prototypical compound for studying fundamental phenomena such as domain formation~\cite{meyer2002ab}, phase transitions, and piezoelectricity~\cite{ghosez2022modeling}.

Over the past decade, several unexpected discoveries in PbTiO$_{\mathrm{3}}$ have significantly advanced our understanding of ferroelectric materials. Notably, distinctive ferroelectric domain walls (DWs) have emerged~\cite{wojdel2014ferroelectric}, redefining their role from passive interfaces to active, dynamic entities with novel functionalities such as heat flow regulation~\cite{langenberg2019ferroelectric}. This paradigm shift establishes DWs as key components for enhancing energy efficiency and reducing power consumption in future electronic devices. Another breakthrough was the prediction of polar skyrmions~\cite{pereira2019theoretical}, topologically non-trivial solitons analogous to their magnetic counterparts. These skyrmions are stable, electrically controllable, and can exhibit intriguing properties such as chirality and tunable negative capacitance~\cite{tikhonov2020controllable,zubko2016negative,guo2025continuously}, positioning them as promising candidates for ultra-dense data storage and advanced memory technologies~\cite{du2024design}.

Understanding such complex structures and phenomena requires computational methods that combine high accuracy with scalability. While density functional theory (DFT) offers reliable atomic-scale insights, its computational cost limits its applicability to large systems or long time scales. Second-principles (SP) interatomic potentials~\cite{Wojdel2013,Escorihuela-Sayalero2017} overcome these constraints by describing the potential energy surface via effective potentials fitted to DFT data~\cite{ghosez2022modeling}. These models have shown predictive capabilities across various studies~\cite{graf2021unified,bastogne2024dynamical,murillo2021coexisting,zhang2023structural,pereira2019theoretical,zubko2016negative,gomez2024switchable,gomez2025inhomogeneous}, advancing the exploration, at the atomic scale, of ferroelectric phenomena of technological relevance.

In this work, we introduce an open-source, fully atomistic second-principles model for PbTiO$_{\mathrm{3}}$, rigorously constructed and validated against DFT data. Our model faithfully captures structural, energetic, and dynamical properties, including phase transitions and finite-temperature phonon softening. We use it to investigate domain-walls and their interactions as well as topological textures such as skyrmions. Doing so, we confirm the usefulness of the approach to uncover previously unexplored polar configurations in PbTiO$_{\mathrm{3}}$.

\section{Methods}
\subsection{First-principles}
We generated the \textit{ab-initio} data for fitting the second-principles interatomic potential using density functional theory calculations with the \textsc{Abinit} software package~\cite{gonze2020abinit}. We used the generalized gradient approximation (GGA) with the PBESol exchange-correlation functional and a planewave-pseudopotential approach, relying on optimized norm-conserving pseudopotentials from the PseudoDojo server~\cite{hamann2013optimized,van2018pseudodojo}. We treated $Pb~5d^{10}6s^26p^2$, $Ti~3s^23p^63d^24s^2$ and $O~2s^22p^4$ electrons as valence states. We adopted a plane-wave energy cutoff of $65\ Ha$ and used a $8\times 8\times 8$ $\Gamma$-centered Monkhorst-Pack~\cite{monkhorst1976special} k-point mesh for the five-atom cubic unit cell, while adapting the k-point sampling to the supercell size in order to maintain a similar high level of convergence. 
We computed the dynamical matrices for the relaxed cubic $Pm\bar{3}m$ on a $4\times 4\times 4$ q-point mesh.
From density functional perturbation theory (DFPT)~\cite{Gonze1997}, we also extracted essential properties, including the dielectric constant, Born effective charges, strain-phonon coupling, and elastic constants. 

\subsection{Second-principles}


The second-principles model employed in this work is designed to accurately and efficiently describe the Born-Oppenheimer potential energy surface (PES) of crystalline materials. It is based on a Taylor expansion of the PES in terms of structural degrees of freedom and around a reference structure—typically a stationary point where all forces and stresses vanish. This reference can represent a configuration relevant to competing phases or a stable structure suited for analyzing local anharmonicity.
The model incorporates all atomic and strain degrees of freedom by systematically distinguishing between harmonic and anharmonic energy contributions. Harmonic interactions, including both short-range forces and long-range dipole-dipole couplings, are derived from DFPT. Anharmonic effects, which capture deviations from harmonic behavior, are introduced through a selection of higher-order terms whose coefficients are fitted to first-principles data, such as energies, forces, and stresses from a representative set of atomic configurations.
Assuming periodic boundary conditions, the model constructs the PES using material-specific parameters. Provided the bond topology remains unchanged, this approach effectively captures the behavior of systems with closely competing structural phases—such as perovskites—where small energy differences drive phase transitions.

We developed the model using the \textsc{Multibinit}~\cite{gonze2020abinit} software package. In practice, the interatomic force constants (IFC), elastic constants, strain-phonon coupling, Born effective charges, and dielectric constant of the cubic phase $Pm\bar{3}m$ were computed using DFPT calculations from \textsc{Abinit}~\cite{gonze2020abinit}. A training set of 3644 DFT configurations was used to select and fit the anharmonic coefficients of the model, applying a cutoff radius of $\sqrt{3}/2$ times the cubic lattice parameter to generate symmetry-adapted terms (SAT) up to the eighth order, including strain-phonon coupling and anharmonic elastic constants. We selected 29 SATs via a goal-function (GF) minimization approach and added 66 additional terms to ensure the boundness of the model energy.

Finite-temperature simulations were performed using a hybrid molecular dynamics–Monte Carlo (HMC) scheme~\cite{duane1987hybrid,betancourt2017conceptual}, as implemented in \textsc{Abinit}~\cite{gonze2020abinit}. Each HMC sweep consisted of 40 molecular dynamics steps with a time step of 0.72 fs, performed in the NPT ensemble. After 500 sweeps for thermalization, simulations typically included 2000 HMC sweeps, increased to 4000 in the vicinity of phase transitions. Structural relaxations used the Broyden-Fletcher-Goldfarb-Shanno (BFGS) algorithm in \textsc{Abinit}~\cite{gonze2020abinit}, and energy barriers were explored using the nudged elastic band (NEB) method with the Quick-Min algorithm~\cite{henkelman2000improved, sheppard2008optimization}. Phonon dispersions at 0~K were obtained with the SP potential in \textsc{Abinit}~\cite{gonze2020abinit}, while finite-temperature phonons were calculated using \textsc{TDEP}~\cite{knoop2024tdep}. This atomistic model has been successfully applied in previous studies~\cite{bastogne2024dynamical,Zatterin2024,gomez2025inhomogeneous,gomez2024switchable}.

\section{Second-principles model}

\subsection{Training set and model construction}
We built the training set (TS) used to fit our second-principles model with the same successful approach used for other compounds such as the prototypical ferroelectric BaTiO$_{\mathrm{3}}$~\cite{zhang2023structural}, non-polar SrTiO$_{\mathrm{3}}$, and antiferroelectric PbZrO$_{\mathrm{3}}$~\cite{zhang2024finite}. By analyzing the phonon dispersion curve, we first identify unstable modes at the $\Gamma$, $M$, and $R$ points, which align with a $2\times 2\times 2$ supercell of the $Pm\bar{3}m$ phase. Condensing then all possible combinations of these unstable modes, we identify six (meta)-stable phases, summarized in Table~\ref{Table:Relax}. These results match closely with previous studies~\cite{sharma2014first}.
\setlength{\tabcolsep}{12pt} 
\begin{table}[H]
\captionsetup{width=\textwidth}
\caption{Structure and energy of various metastable phases of PbTiO$_3$ designated by their space group, as relaxed in DFT. The comparison includes lattice parameters (\AA), distortion amplitudes  (rotations in degrees and total deformation in \AA) respect to the reference cubic phase, and energy gain $\Delta E$ (meV/f.u.) respect to the 5-atom cubic reference structure.}
\begin{center}
\begin{threeparttable}
\begin{tabular}{lcccccc}
\toprule\toprule
\textbf{Phase} & \multicolumn{3}{c}{\textbf{Lattice parameter}} & \multicolumn{2}{c}{\textbf{Distortion}} & $\boldsymbol{\Delta E}$ \textbf{(meV/f.u)}\\
& a (\AA) & b (\AA) & c (\AA) &Rot. ($^\circ$) & Tot.(\AA/f.u)\\
\midrule
$Pm\bar{3}m$ & 3.917  & 3.917     & 3.917        & 0.0  & 0.0     & 0.0    \\
$P4mm$         & 3.861 & 3.861 & 4.244 & 0.0  & 0.423 & -85.24 \\

$Amm2$     & 3.886 & 3.983 & 3.983 & 0.0  & 0.305 & -64.00  \\

$R3m$ & 3.945 & 3.945 & 3.945 & 0.0  & 0.288 & -59.19 \\

$I4/mcm$ & 3.911 & 3.911 & 3.927 & 4.34   & 0.105& -3.27 \\

$P4/mbm$ & 3.915 & 3.915 & 3.920 &2.09 & 0.05 & -0.23 \\

$Imma$ & 3.910 & 3.919 & 3.919 & 3.22 & 0.111 & -3.63\\

$R\bar{3}c$& 3.916 & 3.916 & 3.916 & 2.63 & 0.110 & -3.66 \\

    \bottomrule\bottomrule
\end{tabular}%
    \end{threeparttable}
\end{center}
\label{Table:Relax}
\end{table}
Starting from this information, we build the TS making a smart sampling of configuration space. We explore the PES following linear interpolation paths between all pairs of relaxed phases. Additionally, to capture thermal effects, we add thermal noise to selected configurations by populating phonons from the reference structure~\cite{zacharias2016one}, adjusting the amplitude to ensure that the total energy lies between the energy of the ground state and up to twice that energy above the reference structure. This initial model is then refined: after fitting a preliminary model, we run molecular dynamics simulations at various temperatures using this preliminary model, generating additional data to feed and improve the accuracy of the final model. Combining all data, we end up with a training set that contains 3644 configurations. With these configurations, we have sampled the PES of PbTiO$_{\mathrm{3}}$ around the cubic reference structure, allowing us to fit a final model ready for the validation steps.

\subsection{Validation}
\begin{figure}[h]
\includegraphics[width=\columnwidth]{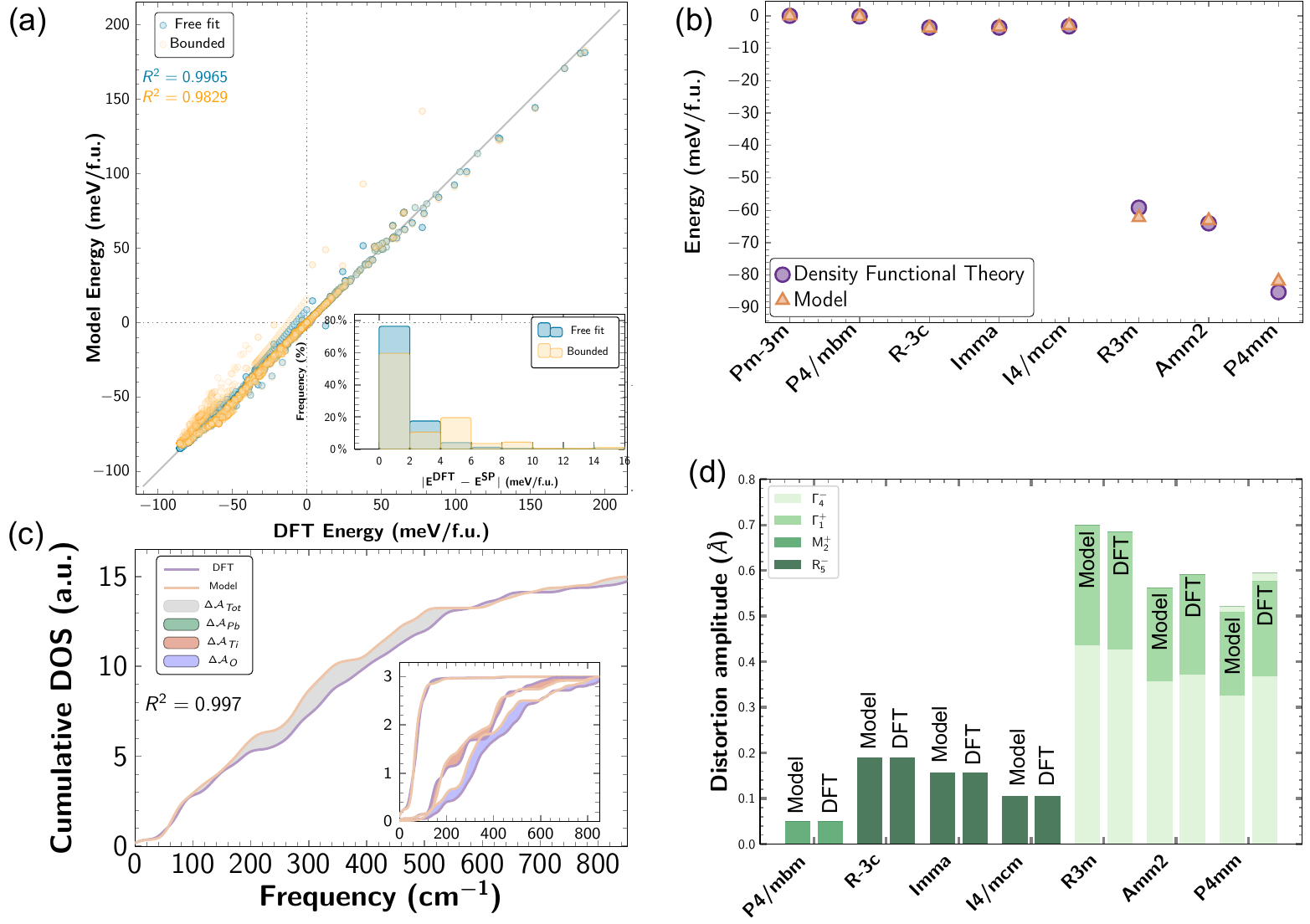}
\caption{Validation against DFT of the interatomic potential for PbTiO$_\mathrm{3}$. (a) Comparison between model and DFT energies, as predicted for the structures contained in the training set. The inset shows the distribution of the amplitude of absolute error made by the model. (b) Comparison between DFT and model energies for different (meta)-stable phases, as consistently relaxed within each approach. (c) Comparison between DFT and model cumulative total phonon density of states in the tetragonal $P4mm$ ground state. The inset shows the comparison between DFT and model contributions to the cumulative phonon density of states individually for each type of atoms. (d) Comparison between DFT and model distortion amplitudes for different meta- and stable phases, as consistently relaxed within each approach.}
\label{fig:validation}
\end{figure}

We first validate our second-principles model by testing its ability to reproduce the data of the training set. As shown in Figure \ref{fig:validation} (a), our free fit model achieves a good fit quality with ($R^2 > 0.996$). The free fit is obtained without any constraint. However, our model consists of a truncated polynomial expansion of the energy around a reference structure. This expansion is often not bounded from below, meaning that the energy can eventually diverge when increasing some displacements of the atoms compared to the reference structure. To address this problem, we performed an automatic bounding of our model by adding higher-order even terms associated with a positive coefficient. As shown in Figure \ref{fig:validation} (a), the bounded model maintains the accuracy of the fit ($R^2 > 0.982$) while eliminating at the same time the possible divergence of the model. 

In order to go one step further in the validation of our model, we performed structural relaxations and compared all stationary phases identified in Figure \ref{fig:validation} (b). As shown in Figure~\ref{fig:validation}(b,c), not only the energies but also the structural distortion obtained from the model-based structural relaxations are highly consistent with the results obtained from the first-principles relaxations. Indeed, the largest error (0.68 meV/atom for the ground state) is within the typical convergence criterion for DFT calculations. Moreover, the biggest difference with respect to the total distortion is equal to 0.049 \AA/f.u., which is negligible when considering the complexity of the system. 

Finally, we computed the cumulated density of states of the phonon dispersion curves in the $P4mm$ phase at 0K using DFPT and our model (see Figure \ref{fig:validation} (c)). To compare them, we calculate the difference of area $\Delta\mathcal{A}$ between the two cumulative density of states (DOS) and normalize it with respect to the DFT result. Then, we obtain a quality factor $\Delta a = 1- \Delta\mathcal{A}/\mathcal{A}_{DFT} = 0.95$ that reflects the quality of our model to reproduce the phonon in this ground state phase. Looking at the cumulated DOS atom by atom, we can attribute the main discrepancy to the oxygen atoms.  

Comparison of the model with an independent DFT test set is also reported in the attached model validation passport \cite{ULG/EW3NNJ_2025}. All together, these results demonstrate that our model can accurately capture not only the PES but also the fine structural details of the system, reinforcing its predictive capability. 

\section{Results}
\subsection{Finite-Temperature Ferroelectric Transition}
\begin{figure}[H]
\includegraphics[width=\columnwidth]{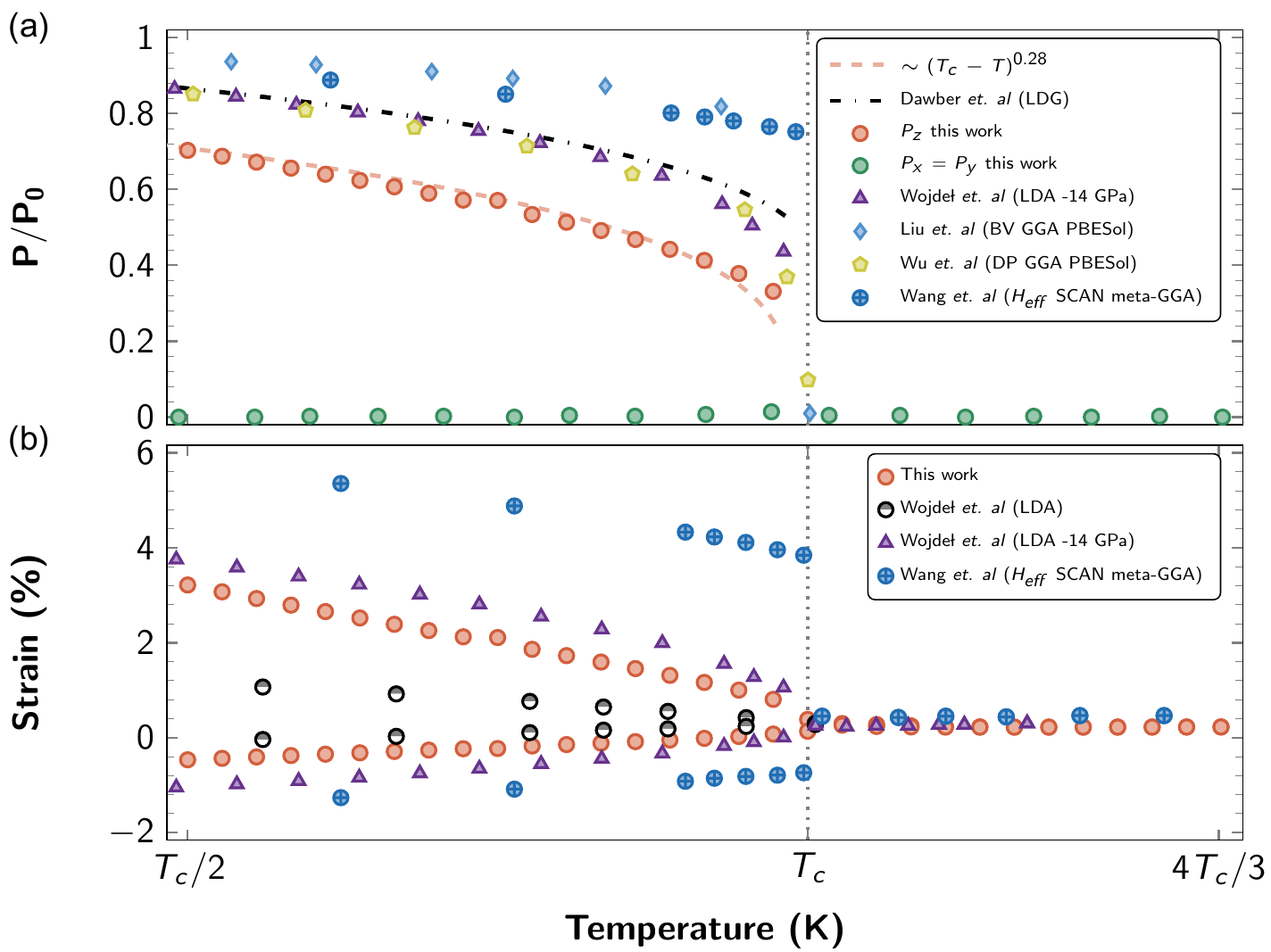}
\caption{Ferroelectric $P4mm$ to paralectric $Pm\mathrm{\bar{3}}m$ phase transition of PbTiO$_\mathrm{3}$ in temperature as predicted  by the interatomic potential . (a) Evolution of the spontaneous polarization (normalized with the polarization $P_0 = 1.1$ C/m$^2$ at 0 K)  with temperature, in comparison with the results of  Refs~\cite{Wojdel2013,liu2013reinterpretation,wu2023modular,dawber2007tailoring,wang2023finite}. The orange dashed curve shows a critical exponent fit of $n=0.28$. (b) Evolution of the homogeneous strains with temperature, in comparison with the results of Refs~\cite{Wojdel2013,xie2022ab}. Note that the data of Ref.~\cite{Wojdel2013} with a negative pressure of -14 GPa have been shifted to align with our strain at $T_c$}
\label{fig:Heating}
\end{figure}

Once the model has been validated, all the advantages of the SP approach can be harnessed, such as large-scale finite-temperature simulations. The first natural finite-temperature simulation to perform concerns the well-known ferroelectric to paraelectric phase transition. As shown in Figure~\ref{fig:Heating}, our model captures the first-order nature of the phase transition but predicts it to occur at 400 K, which is significantly lower than the experimentally observed transition temperature of 760 K~\cite{shirane1951phase}. 

Figure~\ref{fig:Heating} (a) depicts the evolution of the spontaneous polarization $P/P_0$ as a function of temperature, where $P_0$ is s the polarization at 0 K. Our results are compared with theoretical predictions from previous theoretical methods fitted on various data~\cite{Wojdel2013,liu2013reinterpretation,wu2023modular,dawber2007tailoring,wang2023finite}. The polarization decreases with increasing temperature, dropping sharply near the transition temperature, consistent with the first-order character of the transition. 
We notice that the polarization at 0K $P_0$ reported in Figure~\ref{fig:Heating}(a) has been estimated using the Born effective charge of the cubic phase yielding to a slightly overestimated value of 1.1~$C/m^2$. Using instead the Born effective charge of the $P4mm$ phase, would give an underestimated value of 0.84~$C/m^2$. The more accurate $P_0$ calculated for the same structure in DFT using the Berry phase is 0.99~$C/m^2$, which is in good agreement with previous calculations~\cite{yuk2017towards}. 

Figure~\ref{fig:Heating} (b) shows the strain evolution with temperature, highlighting the structural changes accompanying the phase transition. The strain exhibits a discontinuity at $T_c$ further confirming the first-order nature of the transition. Our results are in good agreement with LDA-based calculations under negative pressure (-14 GPa). 

The underestimation of $T_c$ is typical feature of second-principles models,  which is often attributed to the inherent limitations in the underlying exchange-correlation functionals~\cite{zhong1995first,qi2016atomistic,gigli2022thermodynamics} on which the model is fitted. Moreover, here, the high-order development of the model up to order eight combined with the bounding process makes the model stiffer and could contribute to explain the underestimation of the critical temperature of the model. Finally, let us note that the $T_c$ of our model lies between the values obtained from LDA-based second-principles~\cite{Wojdel2013} and machine learning meta-GGA methods~\cite{xie2022ab}, which respectively under- and overestimate $T_c$ in the absence of external pressure. Their closer agreement is only achieved under pressure corection, which our model does not require. Although a weakness, the under- or over-estimation of the critical temperature does not necessarily prevent the practical application of a model and can be handled from simple temperature renormalization as used in the next section.   

\subsection{Temperature-Dependent Phonon Properties}
\begin{figure}[h]
\centering
\includegraphics[width=\columnwidth]{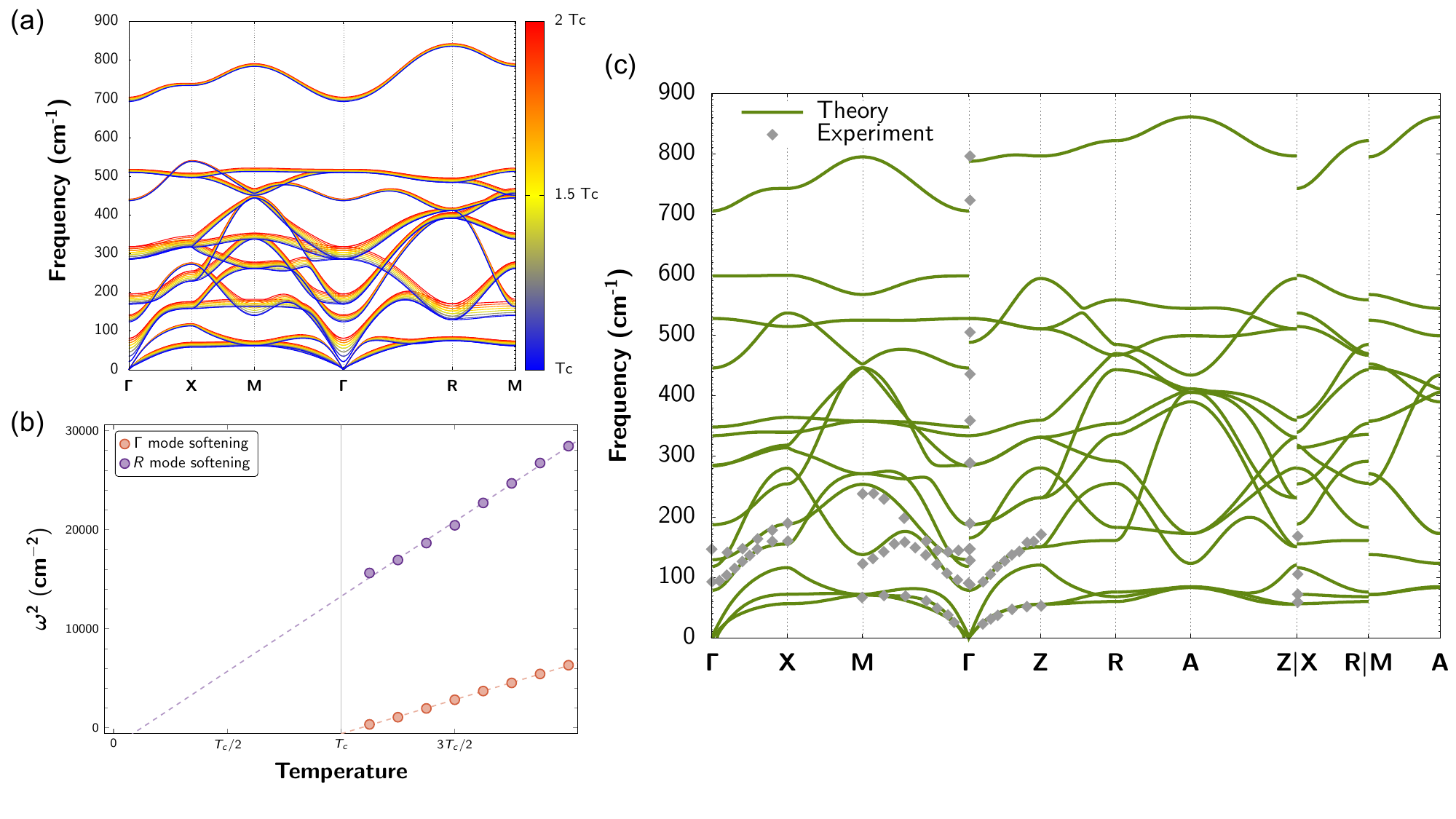}
\caption{\textit{Temperature-dependent phonon dispersion curves of PbTiO$_\mathrm{3}$. (a) Temperature-dependent phonon dispersion in the cubic phase above $T_c$ ($T_c$ stands for critical temperature of the simulation considering $4\times 4\times 4$ supercell). (b) Mode softenings at $\Gamma$ (polar modes) and $R$ (antiferrodistortive modes). (c) Temperature-dependent phonon dispersion curves in the tetragonal phase at $T_{exp}=300\ K$ in a $4\times 4\times 4$ supercell, compared with experimental data from Refs~\cite{tomeno2006lattice,fontana1991raman}.}}
\label{fig:tdep}
\end{figure}

Since PbTiO$_\mathrm{3}$ is known to show a displacive character~\cite{shirane1970soft}, it is interesting to simulate how the phonon dispersion curves are evolving with temperature. For each simulated temperature $T_{sim}$ = $(T_{c,sim}/T_{c,exp})T_{exp}$, we performed a NPT simulation followed by a NVT simulation considering the average volume of the previous simulation. We extract a temperature-dependent dynamical matrix from molecular dynamics and then obtain temperature-dependent phonon dispersion curves. As shown in Figure \ref{fig:tdep} (a), above $T_{c,sim}$, the system is stable in the cubic phase and no unstable phonons have been found. Moreover, a polar mode at $\Gamma$ point is softening from 81 $\mathrm{cm^{-1}}$
to 21 $\mathrm{cm^{-1}}$ when decreasing the temperature from $2\ T_{c,sim}$ to $T_{c,sim}$. Interestingly, other modes are softening at different places throughout the Brillouin zone, in particular at $R$ and $M$.  These antiferrodistortive modes relate to out-of phase (-) and in-phase (+) rotation of the oxygen octahedra. 

As illustrated in Figure~\ref{fig:tdep} (b), $\omega^2$ evolve linearly with temperature in fair agreement with the soft mode theory of Cochran~\cite{cochran1959crystal}. This simultaneous softening at the center and the border of the Brillouin zone is the signature of typical competition between oxygen octahedra rotations and polar distortions in materials with a Goldschmidt tolerance factor close to one such as PbTiO$_{\mathrm{3}}$ and SrTiO$_{\mathrm{3}}$. In the case of PbTiO$_{\mathrm{3}}$, the polar modes are softening faster than the oxygen octahedra rotations, leading to a polar phase transition when the temperature is decreasing, in contrast to the scenario occurring in SrTiO$_{\mathrm{3}}$.

Since our model can predict the temperature-dependent phonon dispersion curve, we can use it to investigate phonons at room temperature ($T_{exp} = 300K$) and compare it with experiment. As shown in Figure \ref{fig:tdep} (c), our model shows excellent agreement with the experimental data, as demonstrated by the close correspondence between the calculated frequencies and the measured frequencies~\cite{tomeno2006lattice,fontana1991raman} over a wide frequency range and between different paths connecting points of high symmetry in the Brillouin zone. This strong concordance highlights the accuracy and reliability of our theoretical predictions in capturing the vibrational properties of the material, despite the underestimation of the critical temperature.

\subsection{Structure and Energetics of Ferroelectric Domain Walls}

Domain walls are extended defects that separate homogeneous regions (domains) with distinct orientations of an order parameter. They represent transitional areas where the order parameter must rotate or vanish, often carrying intrinsic properties that differ significantly from those inside the domains they connect.
In tetragonal PbTiO$_3$, many studies have focused on $180^\circ$ and $90^\circ$ ferroelectric domain walls~\cite{meyer2002ab,poykko1999ab}. The $180^\circ$ walls separate domains of opposite polarizations, while $90^\circ$ walls separate domains with orthogonal polarization directions.

Ferroelectric DWs are classified according to the way the polarization evolves from one domain to the other. If the polarization remains along the same axis and switches by changing its magnitude, the wall is called an Ising wall. In contrast, walls exhibiting a smooth rotation of the polarization within the plane of the wall are called Bloch walls, while those involving a rotation perpendicular to the wall plane are called Néel walls. Originally, DWs in PbTiO$_3$ were considered as Ising walls~\cite{meyer2002ab}. More recently, however, Wojde{\l} and \'{I}\~{n}iguez~\cite{wojdel2014ferroelectric} proposed that 180$^\circ$ DWs can also develop, confined inside the wall itself, an additional ferroelectric component. The latter being oriented perpendicular to the domain polarization, such walls are often called Bloch walls. 
%

Our computations of 180$^o$ DWs structures and energies concern stripe domain configurations in which the domain polarization is oriented along $x$, the domain wall is Pb-centered (confirmed here to be the most stable location ~\cite{meyer2002ab}) and oriented in the $xy$ plane and each domain has a width $t = 10$ u.c. along $z$ ($1 \times 1 \times 20$ simulation box). During the computation, the atomic structure is fully optimized, including simultaneous relaxations of cell parameters and internal degrees of freedom. As explained in Ref. ~\cite{Chege-25}, atomic relaxations can take place along $x$, $y$ and $z$ directions, giving rise to polarization components that can be labelled respectively as Ising, Bloch and Néel. Due to the symmetry breaking produced by the DW, full relaxation of so-called Ising walls typically also naturally includes small atomic relaxations along $z$ (Néel type)~\cite{gu2014flexoelectricity} while, due to its ferroelectric nature, the Bloch component can be present or not. Hereafter, following previous literature, we call ``(relaxed) Ising walls'' those including both Ising and Néel types of relaxations while we refer to those artificially restricted to Ising type relaxation as ``constrained Ising walls''.  We then call Bloch walls those additionally developing a Bloch polarization component along $y$. 

As discussed in Ref. ~\cite{Chege-25}, in stripe configurations, the DW structure and energy only slowly converges with the domain wall density (i.e $1/t$) and the values reported here are therefore specific to $t=10$ u.c.. For 180$^{\circ}$ Ising walls (constrained and relaxed), the DW enegies of the model are comparable to DFT values (Table~\ref{Table:Dw_en}). Also the cell parameters and Bloch and Néel component of the polarization are closely reproduced. As shown in Table~\ref{Table:Dw_en}, the Néel component $P_z$ is much smaller than the Ising and Bloch components, as it originates solely from the strain gradient induced by the domain wall. Nevertheless, despite the small magnitude of $P_z$, the energy gain associated to the atomic relaxation along $z$ is comparable to the energy difference between the fully relaxed Ising and Bloch configurations.

For 90$^{\circ}$ Ising walls,
we followed the same setup as in Ref.~\cite{meyer2002ab}. Table~\ref{Table:Dw_en} shows good agreement between the domain wall energy given by our model and that reported in Ref.~\cite{meyer2002ab}.

\begin{center}
\setlength{\tabcolsep}{16pt} 
\begin{table}[H]
\captionsetup{width=\textwidth}
\caption{Comparison of DW energies (mJ/m$^2$) and maximal component of polarization ($\mu C/cm^2$) for a 180$^\circ$ Pb-centered stripe configuration with a domain width of 10 u.c. (1$\times$1$\times$20 u.c simulation box)  and for a 90$^\circ$ O-O centered with a domain width of 4 u.c. (1$\times$1$\times$8 u.c simulation box). The parenthesis are the comparison with Ref.~\cite{Chege-25} for 180$^\circ$ and Ref.~\cite{meyer2002ab} for 90$^\circ$.}
\begin{center}
\begin{threeparttable}
\begin{tabular}{lccccc}
\toprule\toprule
\textbf{Type} & $\mathbf{E^{DW}}$ & $\mathbf{P_x}$ & $\mathbf{P_y}$  & $\mathbf{P_z}$\\
\midrule
Constraint 180$^\circ$ Ising  &  206 (196)  & 93 (102) & 0.0 (0.0) & 0.0 (0.0) \\
Fully relaxed 180$^\circ$ Ising  & 201 (191)    & 93 (102) & 0.0 (0.0) & 0.15 (0.37)\\
Fully relaxed 180$^\circ$ Bloch  & 195 (187)    & 93 (101) & 36 (36) & 0.17 (0.33)\\
Fixed cell parameters 90$^\circ$ & 39 (32.5)  & 67 & 42 & 3$\times$10$^{-4}$\\
 \bottomrule\bottomrule
\end{tabular}%
    \end{threeparttable}
\end{center}
\label{Table:Dw_en}
\end{table}
\end{center}

\begin{figure}[H]
\centering
\includegraphics[width=\columnwidth]{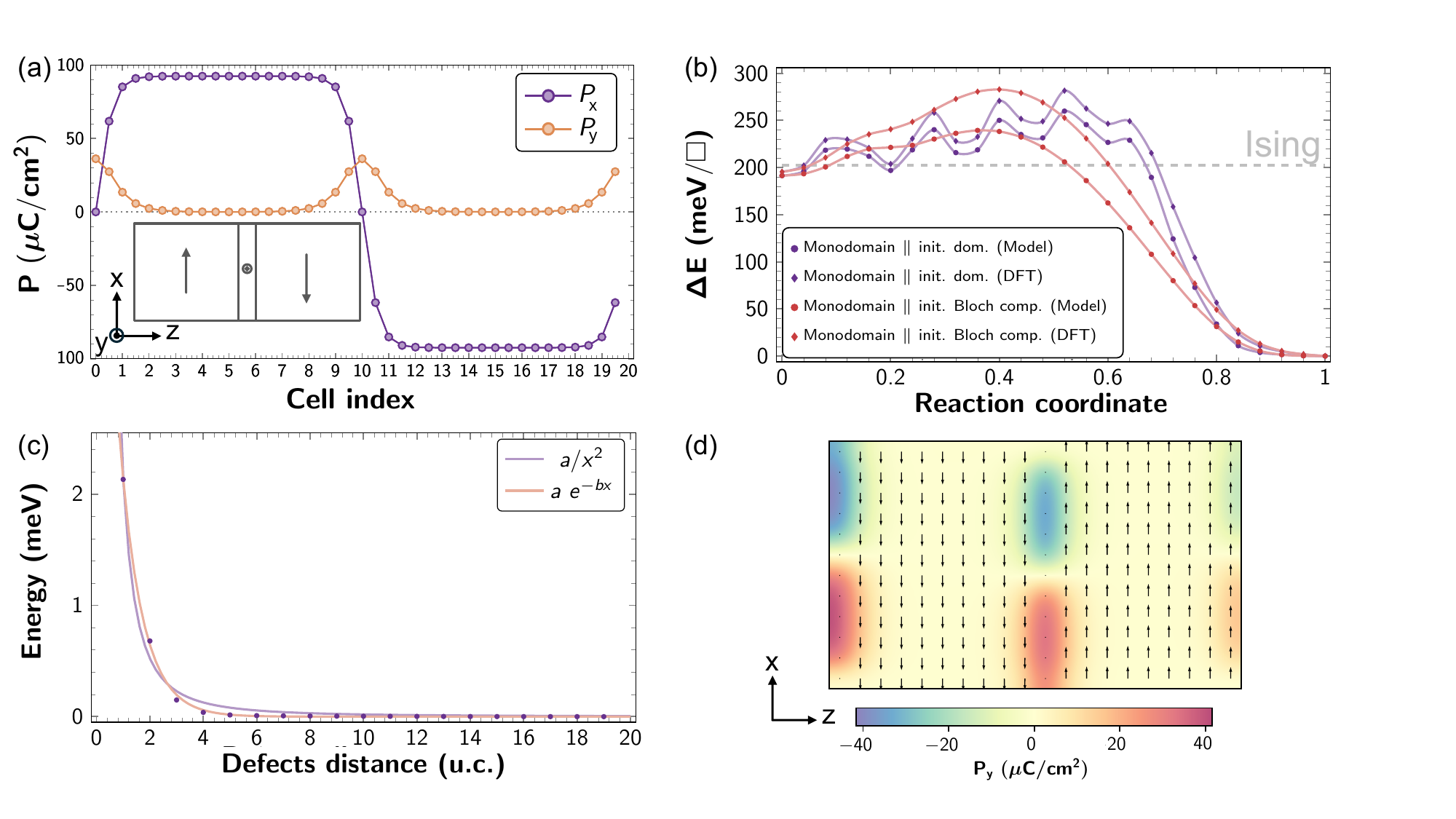}
\caption{\textit{Characterization of 180$^\circ$ Bloch DW in tetragonal PbTiO$_\mathrm{3}$. (a) Polarization profile of $P \parallel [100]$ ($P_x$) (purple) and $P \parallel [010]$ ($P_y$) (orange) in a 180$^\circ$ domain structure. (b) Results of NEB calculations illustrating the domain wall energy evolution along the transition from a Bloch DW to two distinct monodomain configurations: one with the final polarization aligned parallel to the original domain polarization and the other with the final polarization oriented perpendicular to the original domain polarization. Both NEB paths are compared to DFT calculations. (c) Evolution of the energy when DW approach each other and its fit with exponential and power laws. (d) Polarization profile in x-z plane highlighting the appearance of Ising lines within DW.}}
\label{fig:Bloch_DW_sum}
\end{figure}

Our model being validated on conventional Ising $180^\circ$ and $90^\circ$ domain walls, it can now be confidently applied to explore less conventional  configurations like Bloch walls.
Figure \ref{fig:Bloch_DW_sum} (a) presents a striking confirmation of our model's ability to capture the fundamental properties of 180$^\circ$ domain walls (DWs) in PbTiO$_{\mathrm{3}}$ with an additional ferroelectric Bloch component, confined at the domain wall. Specifically, and without having been explicitly trained on such inhomogeneos configurations, it predicts a ferroelectric component within the DW, with a polarization of $P_y = 36~\mu C/cm^2$ (\textit{resp.} $36~\mu C/cm^2$ in DFT), in excellent agreement with previous DFT-based theoretical findings \cite{wojdel2014ferroelectric,wang2014origin,Chege-25} and in much better agreement than the new generation effective hamiltonians~\cite{ma2025active}. Moreover, our calculations reveal an energy difference of approximately $6 \text{ meV/}\square$ between Ising and Bloch DW configurations (for a $1\times 1 \times 20$ supercell, where $\text{meV/}\square$ denotes the energy per unit cell surface area of the DW). This result is in good agreement with DFT (see Table~\ref{Table:Dw_en}), with the energy difference for the whole supercell from DFT and our model differing by 3.8 meV per formula unit. The robustness of our model is further reinforced by its accurate reproduction of the ferroelectric behavior of the Bloch component: the energy of the Bloch DW remains unchanged regardless of whether the Bloch components are aligned or anti-aligned.
These results confirm the deep-seated connection between ferroelectricity and topological structures in PbTiO$_{\mathrm{3}}$, strengthening its role as a cornerstone material for advanced electronics and data storage. The discovery of ferroelectric Bloch-type components confined within DWs \cite{wojdel2014ferroelectric} has revolutionized the field by highlighting their potential in the design of polar skyrmions \cite{pereira2019theoretical} and other emerging topological configurations \cite{chen2020periodic,di2020topology}. Given its extensive range of DW configurations and strong ferroelectric properties \cite{meyer2002ab,poykko1999ab,he2003first}, PbTiO$_{\mathrm{3}}$ continues to be a prime candidate for exploring the interplay between ferroelectricity and topology. Accurately modeling these intricate behaviors is crucial for pushing the frontiers of ferroelectric research and unlocking novel functionalities for future technologies.

To investigate the mechanism by which domain structures with domain walls (DWs) transition to a monodomain state, we performed Nudged Elastic Band (NEB) calculations. This method allows us to systematically determine the minimum energy path for the transition from a configuration with domains to a monodomain state, shedding light on the energetic barriers and intermediate configurations involved in stabilizing a monodomain phase.
A key aspect of this transition is that there are two possible monodomain states: one aligned with the domain and another aligned with the DW in case of Bloch walls. Understanding this process is crucial for predicting how an applied electric field— whether aligned with the polarization in the domain or the DW—will drive the system from a domain structure to a monodomain state. Additionally, this transition provides insight into how DWs vanish with increasing temperature.
Figure \ref{fig:Bloch_DW_sum}(b) presents the NEB calculations performed with our model, showing the predicted transition paths. 

The pathway from a Bloch DW to a monodomain with polarization aligned with one of the original domains involves a DW displacement followed by DW annihilation, leading to several local minima and a rise in energy as the DWs approach each other. Interestingly, the last local minimum arises when the DWs are separated by three unit cells. After that, the DWs are too close to each other and the system evolves continuously to a monodomain state.

In contrast, the transition from a Bloch DW to a monodomain with the polarization aligned with the Bloch component of the DW follows a smooth rotation of the domain.
Both transitions exhibit comparable energy barriers, which are significantly higher than that of the Bloch-to-Ising transition. This suggests that, with increasing temperature, the material will first undergo a Bloch-to-Ising transition before complete suppression of the DW, consistent with previous findings \cite{wojdel2014ferroelectric}.
 Figure \ref{fig:Bloch_DW_sum}(b) also illustrates that our model effectively captures the energetics of the Bloch DW configuration compared to DFT and reproduces the energy path of transition to the monodomain.
 
Furthermore, we investigated how the DWs interact (see Fig. \ref{fig:Bloch_DW_sum} (c)) as they approach each other. Significant DW interactions begin when they are less than three unit cells apart, and their strength increases exponentially as they move closer, which is in good agreement with NEB calculation previously discussed. The interaction is following an approximate trend of $\sim 6.3~e^{-1.1x}$, where $x$ is the distance between DW. For comparison, a fitted power law of the type $2.1/x^{1.8}$  appears less appropriate. 
The exponential decay observed here reflects the short-range interactions between adjacent domain walls. 
This  fast decay contrasts with the more slowly converging behavior (power law) observed in terms of domain-wall density, which is governed by long-range elastic relaxations Ref.~\cite{Chege-25}.

Beyond four unit cells, DWs do not significantly interact, which may lead to disorder of the Bloch component as the temperature rises, complicating its detection. This is due to the ferroelectric nature of the DW, meaning that the relative orientation between two DWs is no longer important beyond four unit cells. 
This transition behavior highlights the versatility of our model in capturing complex domain dynamics and the intricate energy landscapes of PbTiO$_{\mathrm{3}}$. 

To deepen our understanding of DW structures, we examine potential disorders that may arise within them, focusing on the stabilization of Ising lines, as proposed in BaTiO$_{\mathrm{3}}$~\cite{stepkova2015ising}. Figure \ref{fig:Bloch_DW_sum} (d) shows that the relaxed structure inside the 180$^\circ$ DW of PbTiO$_{\mathrm{3}}$ closely resembles the configuration predicted by Stepkova et al. for 180$^\circ$ Bloch DWs in BaTiO$_{\mathrm{3}}$~\cite{stepkova2017possible} and observed experimentally in LiTaO$_{\mathrm{3}}$~\cite{cherifi2017non} . In our case, the Ising lines configuration is 2.95 $\text{meV/}\square$ higher than the aligned Bloch configuration. To the best our knowledge, this is the first report of Ising lines in PbTiO$_{\mathrm{3}}$, which call for experimental verification. Independently, this also highlights the challenge of detecting Bloch components using experimental techniques such as XRD, which require a well-ordered Bloch component for accurate observation~\cite{Zatterin2024}.


\subsection{Stabilization of Polar Skyrmionic Textures}

\begin{figure}[H]
\centering
\includegraphics[width=\columnwidth]{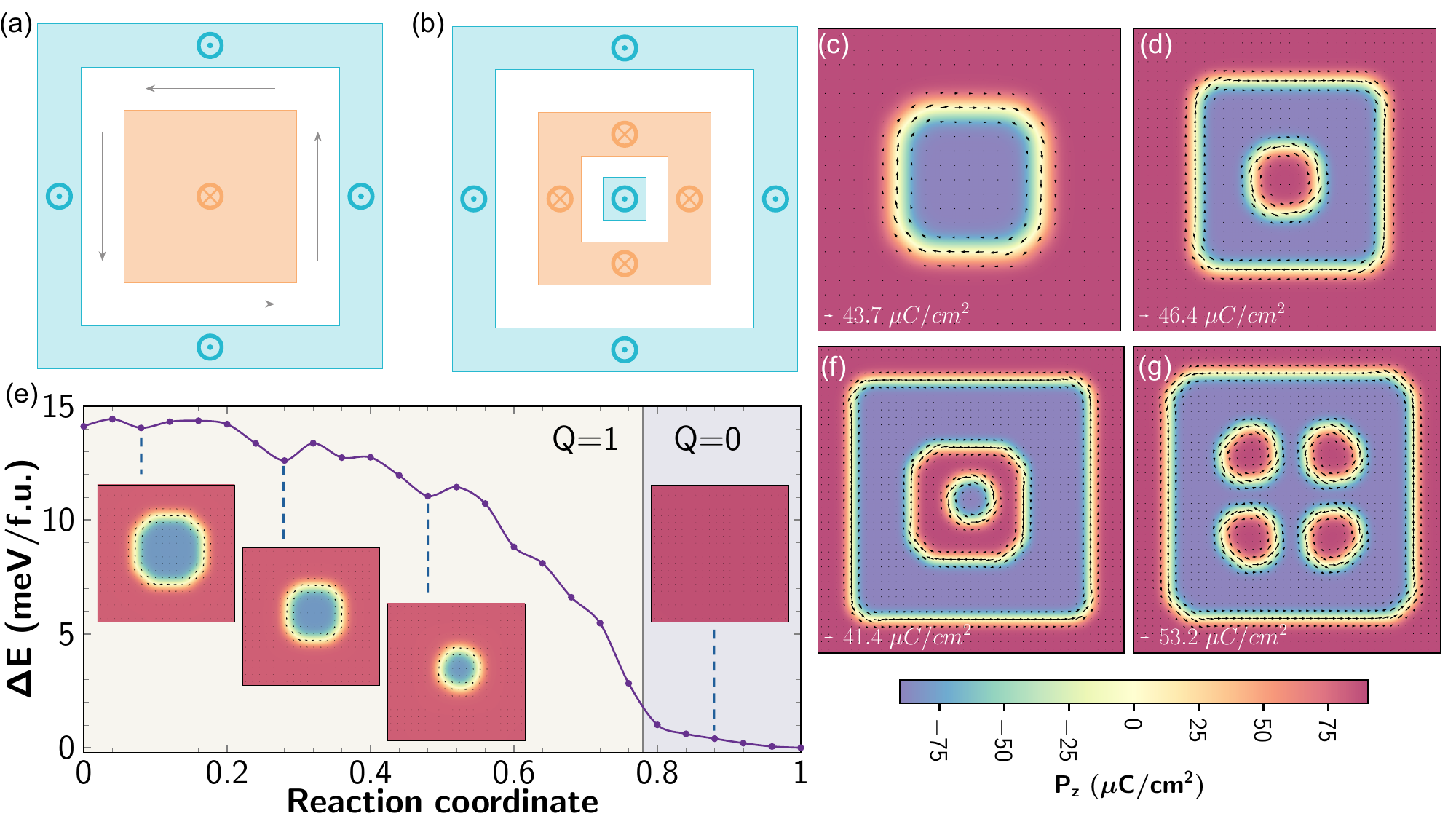}
\caption{\textit{Skyrmion, skyrmionium, target-skyrmion and skyrmion bag in PbTiO$_\mathrm{3}$. (a) Sketch of a Bloch skyrmion. (b) Sketch of a skyrmionium constructed using $180^{\circ} $ domains. (c)  Relaxed Bloch skyrmion using a $20\times 20\times 1$ supercell, starting from the configuration of panel (a). (d) Relaxed Bloch skyrmionium using a $32\times 32\times 1$ supercell, starting from the configuration of panel (b). (e) Results of a NEB calculation along the path between the skyrmion and monodomain states using a $20\times 20\times 1$ supercell. (f) Relaxed Bloch target-skyrmion using a $41\times 41\times 1$ supercell. (g) Relaxed Bloch skyrmion bag using a $40\times 40\times 1$ supercell.}}
\label{fig:Bloch_skyrmion_sum}
\end{figure}

Since our model naturally reproduces the Bloch DW, which forms the fundamental structure of the Bloch skyrmion in PbTiO$_{\mathrm{3}}$, it is natural that we can stabilize this configuration. 
Skyrmions are topological textures characterized by a rotation of an internal order parameter (spins, polarization, ...), forming vortex-like structures with a well-defined topological charge. They emerge as stable or metastable configurations in systems where competing interactions or boundary conditions give rise to nontrivial winding patterns.
Figure \ref{fig:Bloch_skyrmion_sum} shows the stabilization of a polar Bloch skyrmion using our model, following the procedure described in Ref.~\cite{pereira2019theoretical}. The polarization profile exhibits a tightly bound, vortex-like arrangement, with polarization components forming close patterns around a central core that is polarized in the opposite direction of the surrounding matrix. Different recent theoretical studies further theoretically demonstrated that acoustic phonon excitation (APEX)~\cite{bastogne2024dynamical} or an inhomogeneous electric field~\cite{gomez2025inhomogeneous} can dynamically stabilize such skyrmions, allowing controlled creation and manipulation of these topological structures.

To analyze the transition from a skyrmion to a monodomain state, we perform a NEB calculation. Figure~\ref{fig:Bloch_skyrmion_sum}(e) illustrates the minimum energy pathway, where the skyrmion progressively shrinks while transforming into a monodomain structure. Along this pathway, several intermediate configurations are observed, some of which correspond to metastable states with distinct local energy minima. 
This finding agrees with finite-temperature simulations. Indeed, the skyrmion remains stable up to 50~K in a $20\times 20\times 4$ supercell. Beyond this temperature, thermal effects induce its collapse into a monodomain state through a pinning–depinning transition. This behavior aligns with the observations of Liu \textit{et al.} in Ref.~\cite{liu2016intrinsic}. 

In addition, the last minimum exhibited by the NEB calculation corresponds to a skyrmion radius of four unit cells, which is consistent with the domain wall interaction length discussed in the previous section.

Given the robustness of our model in reproducing topologies reported in the literature, we extend our study to explore additional polar textures.

Figure~\ref{fig:Bloch_skyrmion_sum}(d) shows a Bloch-type skyrmionium stabilized in PbTiO$_3$. A ``skyrmonium'' is a complex polar texture characterized by concentric polarization domains. In this configuration, the central core of given polarization is encircled by a ring of opposite polarization, forming a nontrivial spin structure with net zero topological charge. Analogous textures have been previously reported in ferromagnetic systems under various names, including "2$\pi$-vortex"~\cite{bogdanov1999stability,hagemeister2018controlled} and "donut skyrmion"~\cite{streubel2015manipulating}, the term skyrmionium is usually given to a skyrmion surrounded by a domain wall of opposite topological charge~\cite{gobel2021beyond} and it is the convention we are adopting here. In magnetic systems, the relatively simple stabilization of skyrmioniums has made them promising candidates for non-volatile memory applications~\cite{kolesnikov2018skyrmionium}.

The term "target-skyrmion"~\cite{zheng2017direct} is also sometimes  used for such structures but, rigorously speaking, target-skyrmion more precisely refers to multi-ring textures with nonzero topological charge. In our vocabulary, the target-skyrmion so consists of multiple concentric polarization domains, giving rise to a nonzero topological charge. They were also successfully stabilized in PbTiO$_3$, as illustrated in Figure ~\ref{fig:Bloch_skyrmion_sum}(d).

Finally ``skyrmion bags'' are also sometimes reported in magnetic systems~\cite{zheng2017direct,foster2019two}. They consists in several skyrmion-like cores confined within a larger enclosing domain wall, resulting in a composite texture with higher total topological charge. They can also be stabilized in PbTiO$_3$ as illustrated in Figure ~\ref{fig:Bloch_skyrmion_sum}(f).

Figure ~\ref{fig:Bloch_skyrmion_sum}  highlights the rich diversity of topological structures accessible in ferroelectrics and point to new possibilities for information encoding based on topological textures.
These structures can be understood within a unified $k\pi$-skyrmion framework, where the out-of-plane polarization component reverses $k$ times between the center and the periphery. Within this scheme, the standard skyrmion, skyrmionium, and target-skyrmion correspond to $k=1$, $k=2$, and $k=3$, respectively, reflecting increasing radial complexity and domain wall multiplicity~\cite{bogdanov1999stability}.

In our case, these textures are stabilized by selectively reversing the $P_z$ component within the skyrmion core. By controlling the number and spatial arrangement of these reversals, we can in principle realize $k\pi$-skyrmions with arbitrary values of $k$, constrained only by the physical size of the system and the spatial resolution of the applied electric field.

While skyrmionium textures have been observed in PbTiO$_3$/SrTiO$_3$ superlattices, where interfacial coupling facilitates their formation~\cite{kasai2024mechanical,linker2022squishing}, our results demonstrate that a broader variety of topological structures can also be stabilized in bulk PbTiO$_3$.
 This has been achieved here by intentionally imposing a guess atomic structure for the distinct textures and subsequently relaxing the original configuration but we have checked that these various textures can also be stabilized via the activation of appropriate localized Gaussian electric field as proposed  in Ref.~\cite{gomez2025inhomogeneous}, which further opens the way to their dynamical manipulation.

\subsection{Influence of Biaxial Strain on Topological Defects}

\begin{figure}[H]
\centering
\includegraphics[width=\columnwidth]{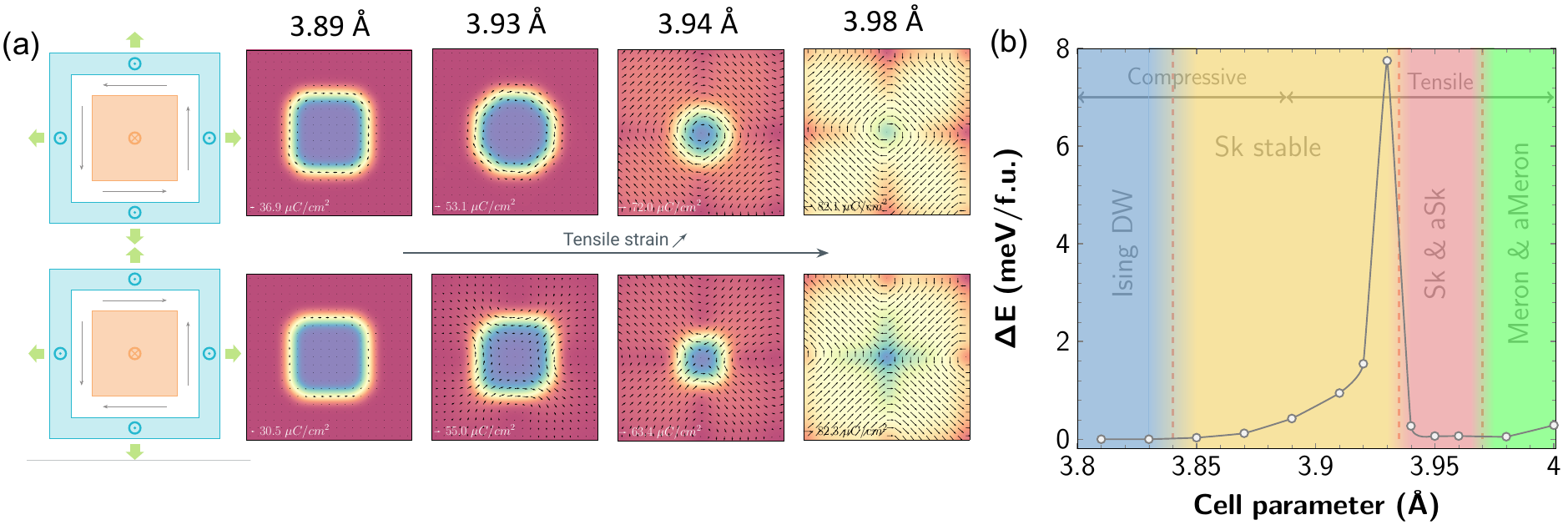}
\caption{\textit{Effect of biaxial strain on the stability of skyrmion and anti-skyrmion in PbTiO$_\mathrm{3}$. (a) Skecth a the initial configuration of a skyrmion and a anti-skyrmion and the evolution of the relaxed polar texture of skyrmion and anti-skyrmion under biaxial strain.  (b) Evolution of the difference of energy between skyrmion and anti-skyrmion for various values of in-plane cell parameters.}}
\label{fig:Bloch_Askyrmion_sum}
\end{figure}

An anti-skyrmion is a topological texture characterized by a  vorticity opposite to that of a skyrmion. It typically involves an inverted internal configuration. Skyrmions exhibit a characteristic rotational sense—clockwise or counterclockwise. In contrast, anti-skyrmions display the opposite winding. The skyrmions discussed earlier emerge from the Bloch component of the DW. An anti-skyrmion can thus be constructed by reversing the internal structure of this component, as shown in Fig.~\ref{fig:Bloch_Askyrmion_sum}(a). This approach is justified by the ferroelectric nature of the Bloch component.

In the absence of strain constraints, the skyrmion is more stable than the anti-skyrmion, as shown in Fig.~\ref{fig:Bloch_Askyrmion_sum}(b). The anti-skyrmion is not stable at finite temperature. Indeed, it transforms into a skyrmion even at low temperatures. Under epitaxial tensile strain, the energy difference between the skyrmion and the anti-skyrmion increases, as shown in Fig.~\ref{fig:Bloch_Askyrmion_sum}(b). This is due to the enhancement of the in-plane component under tensile strain. It increases the energetic cost of the head-to-head configuration in the anti-skyrmion. However, at a critical threshold (around 3.94~\AA), the energy difference abruptly drops to zero. This effect is analogous to the recent theoretical prediction of [001]$_\mathrm{pc}$ skyrmions and anti-skyrmions in the rhombohedral $R3m$ phase of BaTiO$_3$ and KNbO$_3$~\cite{gomez2024switchable}. 

For cell parameters above 3.94~\AA, the in-plane component extends throughout the supercell. As a result, skyrmions and anti-skyrmions become energetically degenerate. This is due to periodic boundary conditions and the Poincaré–Hopf theorem~\cite{Milnor-65} which dictates that the total vorticity in the simulation cell must vanish. As a result, vortex and antivortex textures must cancel each other.
It is worth noting that the predicted skyrmion structures in BaTiO$_3$, KNbO$_3$, and those generated by our model closely resemble the skyrmions previously reported in PbTiO$_3$ under tensile strain~\cite{pereira2019theoretical}, although anti-skyrmions have not yet been observed in PbTiO$_3$ to date. Finally, Fig.~\ref{fig:Bloch_Askyrmion_sum}(b) also shows that at high compressive epitaxial strain, the energy difference between skyrmions and anti-skyrmions tends toward zero. 

Under compressive strain, the Bloch component diminishes. As a result, both textures approach an Ising-type domain wall. At high tensile strain, the out-of-plane component localizes at the vortex and antivortex cores. This transforms skyrmions and anti-skyrmions into merons and anti-merons.
Interestingly, Gómez-Ortiz \textit{et al.}~\cite{gomez2024switchable} also demonstrated that skyrmions and anti-skyrmions can be stabilized through the application of a spatially modulated electric field~\cite{gomez2024switchable}. A similar approach could be envisioned here to stabilize skyrmions and antiskyrmions but also merons and anti-merons.

In this section, we have shown that our model effectively captures the formation, stability, and transformation of defects in PbTiO$_{\mathrm{3}}$, including DWs, skyrmions, antiskyrmion, meron, anti-meron and skyrmionium. This capability offers a comprehensive understanding of defect dynamics and reinforces the model’s reliability in representing complex ferroelectric behavior.
These achievements not only expands the known polar topologies in PbTiO$_{\mathrm{3}}$ but also highlights its potential for hosting diverse topological structures under strain.

\section{Conclusions}

We have developed an atomistic second-principles model for PbTiO$_{\mathrm{3}}$ that successfully reproduces critical physical phenomena, including phase transitions, temperature-dependent phonon behavior, and complex DW dynamics and allow the discovery of new polar textures in PbTiO$_{\mathrm{3}}$ such as Ising lines, skyrmionium, target-skyrmion, skyrmion bag, antiskyrmion, meron and anti-meron. By providing a genuinely open-source, user-friendly implementation compatible with \textsc{Multibinit}~\cite{gonze2020abinit}, we aim to bridge the gap between small-scale DFT calculations and large-scale simulations essential for practical applications. This work aligns with the principles of open science, ensuring that the model is accessible and reproducible, thereby encouraging collaboration and innovation in the study of ferroelectric materials. We hope this contribution serves as a foundation for further advances in materials research, fostering a transparent and cooperative scientific community.

\section*{Data availability}
All data are available in the main text. The second-principles model is open-source and available at Ref~\cite{ULG/EW3NNJ_2025} along with a validation assessment.
\acknowledgments
The authors gratefully acknowledge Fernando Gómez-Ortiz for his careful reading of the manuscript and for his constructive advice and insightful discussions. They also thank Bertrand Dupé for valuable discussions. This work was supported by the European Union’s Horizon 2020 research and innovation program under grant agreement number 964931 (TSAR) and by F.R.S.-FNRS Belgium under PDR grants T.0107.20 (PROMOSPAN) and T.0128.25 (TOPOTEX). 
The authors also acknowledge the use of the CECI supercomputer facilities funded by the F.R.S-FNRS (Grant No. 2.5020.1) and of the Tier-1 supercomputer of the Fédération Wallonie-Bruxelles funded by the Walloon Region (Grant No. 1117545). 
\newpage
\medskip
\bibliography{reference}

\end{document}